\documentclass[11pt,twoside]{article}
\usepackage{./asp2014}

\aspSuppressVolSlug
\resetcounters

\begin{document}

\title{Science with a ngVLA: Imaging planetary systems in the act of forming with the ngVLA}

\author{Luca Ricci \vspace{1mm}
\affil{Department of Physics and Astronomy, California State University Northridge, 18111 Nordhoff Street, Northridge, CA 91330, USA; \email{luca.ricci@csun.edu}}
}
\author{Andrea Isella, Shang-Fei Liu \vspace{1mm}
\affil{Department of Physics and Astronomy, Rice University, 6100 Main Street, Houston, TX 77005, USA}
}
\author{Hui Li \vspace{1mm}
\affil{Theoretical Division, Los Alamos National Laboratory, Los Alamos, NM 87545, USA}
}

\paperauthor{Luca Ricci}{luca.ricci@csun.edu}{}{California State University}{}{Northridge}{CA}{91330}{USA}

\begin{abstract}
The discovery of thousands of exoplanets has shown that the birth of planets is a very efficient process in nature. Several physical mechanisms have been proposed to describe the assembly of planets in disks surrounding young stars. However, observational constraints have been sparse on account of insufficient sensitivity and resolution, especially for imaging the inner $10 - 20$ au from the star, where most of planets are observed. 
Thanks to its unprecedented angular resolution and sensitivity at wavelengths where the emission from the circumstellar material is optically thin, the ngVLA has the potential to transform our understanding of planet formation. In this chapter, we present state-of-the-art theoretical models of planetary systems in the act of forming that demonstrate ngVLA capabilities of imaging and follow the temporal evolution of young solar system analogues up to the distance of the Orion Nebula. These images will unveil how planets form and interact with their parental disks, and will shed light on the diverse properties of exoplanetary systems. 


\end{abstract}

\section{Introduction}
The discovery of thousands of exoplanets over the last couple of decades is a clear sign that the birth of planets is a very efficient process in nature \citep[e.g.,][]{Burke:2015}. Theories invoke several possible mechanisms to describe the assembly of planets in the disks around pre-main-sequence stars, but observational constraints have been sparse on account of insufficient sensitivity and resolution. Understanding how planets form and interact with the disk is crucial also to illuminate the main characteristics of a large portion of the full planet population that is inaccessible to current and near-future observations. 

Perhaps the most puzzling characteristic of the demographics of exoplanets is the broad range of observed orbital radii, which span from a fraction to several tens of au. In particular, the discovery of several ``hot Jupiters'' has indicated that the orbits of planets can significantly vary right after their formation \citep{Mayor:1995}. Indeed, theory indicates that forming planets can easily migrate either towards or away from the star through exchange of angular momentum with the parent circumstellar material \citep{Baruteau:2014}.
Furthermore, once the parent disk has dissipated, planets can change orbital radii due to the interaction with other planets in the same system.
This means that, most likely, the orbital parameters of mature exoplanets do not correspond to the initial location of their formation. Knowing where planets of different masses form in the disk is crucial to constrain the dynamical, physical and chemical history of planetary systems.

Only recently, infrared and (sub-)millimeter telescopes have achieved the angular resolution
required to spatially resolve the inner regions of nearby proto-planetary disks, down to distances of $\sim 10 - 20$ au from the central star. 
Recent high angular resolution observations at these wavelengths resulted in the discovery
of morphological features, e.g., rings and spirals, in the distribution of circumstellar gas and dust with characteristic sizes larger than 20 au \citep[e.g.,][]{Casassus:2013,ALMA:2015,Isella:2016,Dipierro:2018}. 
These structures are thought to originate from gravitational perturbations by yet unseen planets in the act of forming, and provide a tool to measure planet masses and orbital radii \citep{Jin:2016, Liu:2018}.
However, these observations also showed that the disk emission within the inner $\sim 10 - 30$ au is typically optically thick at sub-millimeter and shorter wavelengths \citep[e.g.,][]{Carrasco-Gonzalez:2016}. This makes very difficult, if not impossible, to detect density perturbations induced by planets in the disk regions where most of the planets are expected to form.
The solution to this problem is to image protoplanetary disks at wavelengths longer than about 3~mm, where even the dust emission from the innermost disk regions is optically thin \citep[e.g.,][]{Testi:2014}. However, existing telescopes operating at these wavelengths such as the Jansky Vary Large Array (JVLA) and the Australia Telescope Compact Array (ATCA) do not achieve the angular resolution nor the sensitivity required to resolve the innermost disk regions.

The unprecedented angular resolution and sensitivity of a next-generation Very Large Array (ngVLA), as currently proposed, would revolutionize our understanding of planet formation by enabling routine observations of the inner $\sim 10-20$ au from the star at spatial resolutions below 1 au. ngVLA observations have the potential to reveal forming planets with masses as low as few Earth masses and  unveil the dynamics of solid particles as they interact to form planetesimals. 
Coupled with its nearly order of magnitude increased sensitivity and angular resolution compared to the current JVLA, the ngVLA makes it possible to investigate the processes of planet formation and disk-planet interaction for large, heterogeneous samples of disks orbiting stars with different properties.
Furthermore, by taking advantage of the relatively short orbital periods of planets at $< 10$ au from Solar-mass stars as well as by its high astrometric precision, the ngVLA can measure the motion of disk substructures as they orbit around the central star on timescales as short as a month.
Finally, by imaging and monitoring the temporal evolution of young solar system analogues, the ngVLA would directly inform about the origin of our own planetary system. 


In this chapter, we focus on how the ngVLA can be used to map the dust continuum emission in the inner regions of circumstellar disks and reveal the presence of 
young planets with different masses and orbital radii. Our main goal is to show what planets can generate disk substructures that would be directly detectable with an instrument like the ngVLA in its current proposed design. 



\section{ngVLA observations of planets in the act of forming}

To investigate the capabilities of the ngVLA in 
detecting and characterizing planetary systems in the act of forming, 
we started by performing numerical simulations using the 2D LA-COMPASS (Los Alamos COMPutional Astrophysics Simulation Suite) bi-fluid hydrodynamic
code \citep{Li:2005,Fu:2014}. This code, which accounts for the dynamics of gas, dust and their mutual interaction, was used to model the gravitational interaction between planets with different masses and orbital radii and gas and dust in the disk. The output of the hydrodynamic simulations were then coupled to the Monte Carlo Radiative Transfer code RADMC3D \citep{Dullemond:2012} to generate synthetic multi-wavelength images of disks perturbed by planets, that were finally used to simulate ngVLA observations of the dust continuum at wavelengths of about 3 mm and 1 cm. 
As a term of comparison, we also simulated ALMA observations for the same models. This was to investigate the potential improvement allowed by the ngVLA over the observational information that can be gathered for the same disk with ALMA, as well as the possible complementarity of the two instruments, given that ALMA is particularly sensitive to small-scale disk structures at wavelengths shorter than 3 mm. 
The details of the method and main assumptions are described in \citet{Ricci:2018}, while the main results are summarized below.

\subsection{Revealing planets in the terrestrial planet forming region}

ngVLA observations that achieve an angular resolution of 5 mas at a wavelength of 3 mm can detect and characterize disk structures generated by planets within 10 au from young Solar-like stars in nearby star forming regions. At a distance of 140 pc, an angular resolution of 5 mas corresponds to a spatial resolution of 0.7 au.

In particular, Figure~\ref{fig:ngvla_planets} shows that the ngVLA can detect and spatially resolve strong azimuthal asymmetries in the disk generated by the interaction with planets as massive as Saturn or heavier. These pronounced asymmetries in the dust continuum emission are caused by the trapping of dust particles towards the local maximum of the gas density and pressure, the asymmetric structure of the gas being shaped by the Rossby Wave Instabilities triggered by the gap opened by the planet gravitational torques~\citep{Lovelace:1999,Li:2000,Lyra:2009}. 

Furthermore, the ngVLA is capable of detecting azimuthally symmetric rings and gaps generated by planets with masses below the mass of Neptune of $17~M_{\oplus}$.
The disk models shown in Figure~\ref{fig:ngvla_planets} assume a value of $10^{-5}$ for the $\alpha$-parameter of the gas viscosity. \citet{Ricci:2018} have also shown that in the case of a more viscous disk, with $\alpha = 10^{-3}$, the ngVLA would still resolve gaps opened by planets with masses as low as a few tens times the mass of the Earth. 

\articlefigure{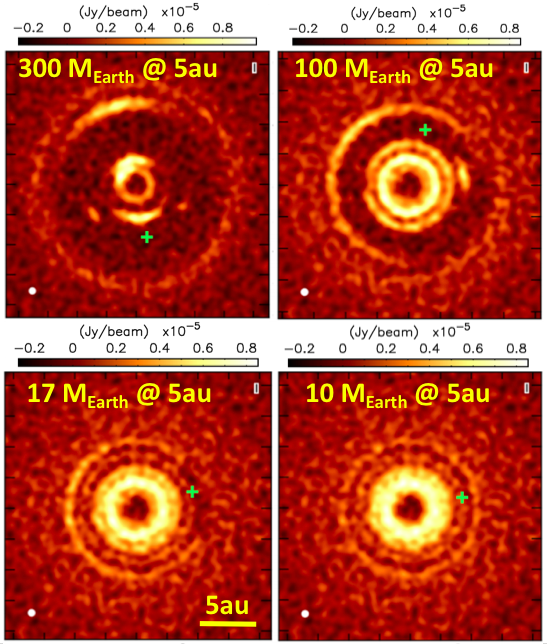}{fig:ngvla_planets}{Simulations of ngVLA observations for the dust continuum at a wavelength of 3 mm for a disk with planets at 5 au from the central star. The planet masses are labeled in each panel, and the planet locations are indicated by the green crosses. The white ellipses show the size of the
synthesized beam, i.e. 5 milliarcsec, corresponding to 0.7 au at the assumed distance of 140 pc. In terms of the adopted sensitivity, the rms noise level for these observations is 0.5 $\mu$Jy/beam~\citep[adapted from][]{Ricci:2018}.}

\subsection{Studying the formation of Super-Earths}

In \cite{Ricci:2018}, we have investigated ngVLA capabilities of detecting planet perturbations by varying the main physical parameters of the disk. For example, it is expected that the opening of a gap at the orbital radius of the planet depends on the planet mass and orbital radius, as well as on the disk viscosity and gas pressure forces, as they both act against the mechanism of gap opening by the planet's gravity \citep[e.g.,][]{Goldreich:1980}. Theory predics that disks characterized by low viscosity and weak gas pressure forces, the latter corresponding to low vertical pressure scale height, will be prone to gap opening by planets with lower mass than in a disk with higher viscosity and stronger pressure forces. 

By varying the level of the viscosity parameter $\alpha$ and the pressure scale height $h$, we found that ngVLA observations could reveal the presence of planets with masses as low as $\approx 5~M_{\oplus}$ at an orbital radius of 5 au. Such observations would provide direct information on the origin of super-Earths, the most common among the different types of planets detected so far, and on the environment in which they form.



\subsection{Studying planet formation in diverse environments}

As discussed above, ALMA observations are capable of probing the formation of planets orbiting at more than 10-20 au from their parent star in nearby low mass star forming regions such as, for example, Taurus and Ophiuchus. Whereas this provides a huge improvement to what has been possible in the past, these star forming regions cannot be considered as representative of star and planet formation. Most of stars and planets form indeed in massive star forming regions, where both the harsh radiation environment and the higher stellar density are expected to influence the chemistry and architecture of planetary systems. The Solar System itself was though to have formed in a much richer cluster not comparable to nearby star forming regions \citep{Adams:2010}.

Based on existing study of circumstellar disks, \citet{Ricci:2018} have estimated that ngVLA observations with similar, or higher, signal-to-noise ratios, than those presented in Figure 1 are achievable for at least a couple hundreds disks within 400 pc from the Sun, a region that would encompass a large fraction of the Orion Molecular Complex, including the Orion Nebula Cluster \citep{Robberto:2013}. An even larger sample could be obtained by extending these observations to disks in more massive star forming regions, such as Cepheus OB3, at 600$-$800 pc. Although the spatial resolution achievable at these 
distances would be a factor of several lower than for closer disks, ngVLA observations would still be able to clearly detect gaps opened by giant planets at orbital radii of about 5 au, and larger. 

Therefore, thanks to its high angular resolution and sensitivity, the ngVLA would be capable of studying planet formation in disks in very different astrophysical environments. Besides probing environments that are more representative of star and planet formation than those sampled by the closest young regions, this investigation would allow us to quantify the effects of high stellar density and high energy radiation field on the potential that a disk has to form planets, thus providing critical inputs to theories of planet formation and early evolution.

\subsection{The case of the young Solar Nebula}

As a special case for our disk models, we investigated ngVLA capabilities of probing planetary systems in formation similar to the young Solar nebula.
To this end, we performed a simulation of a disk harboring a multiple planets with masse and orbital radii as in the Solar nebula model by \citet{Walsh:2011}. This model, which was proposed to explain a number of observational facts in our Solar System, predicts that after a few million years from the formation of the Sun, Jupiter, Saturn, Neptune and Neptune were located at orbital radii of about 3.5, 4.5, 6 and 8 au from the early Sun, respectively.  

\articlefigure{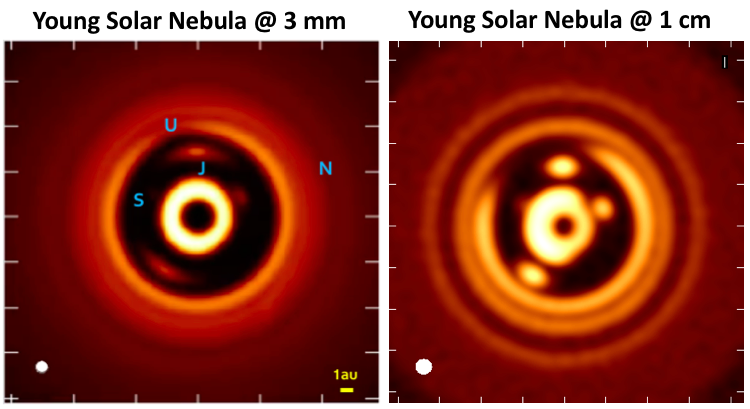}{fig:ngvla_walsh}{Left panel) Simulation of ngVLA observations for the dust continuum at a wavelength of 3 mm for a disk with planets at about 3.5 (Jupiter), 4.5 (Saturn), 6 (Uranus) and 8 au (Neptune) from the central star. The letters indicate the location of each planet. The white ellipse in the bottom left corner shows the size of the
synthesized beam, i.e. 5 milliarcsec, corresponding to 0.7 au at the assumed distance of 140 pc. The rms noise level for these observations is the same as in Fig.~\ref{fig:ngvla_planets}. The reason for the higher signal-to-noise ratio than in Fig.~\ref{fig:ngvla_planets} are the assumed higher gas and dust densities, a factor of $\approx 10\times$ higher than for the models in the previous figure. Right) Same as for left panel, but for ngVLA observations at 1 cm. In this case, the surface brightness of the disk model has been rescaled up to produce a total disk flux of $\approx 1$ mJy at 1 cm, similar to what estimated for the dust emission at this wavelength in a small sample of bright disks in Taurus \citep{Rodmann:2006}. The size of the synthesized beam is 7 mas, or 1 au at 140 pc, and the assumed noise on the map is about 70 nJy$/$beam.}     

Figure~\ref{fig:ngvla_walsh} shows the results of our ngVLA simulated continuum observations at a wavelength of 3 mm (left panel) and 1 cm (right). A large gap is opened by the young Jupiter and Saturn between disk radii of $\approx 3$ and 5 au. Smaller individual gaps are opened by Uranus and Neptune, and the are both detected with the ngVLA. Inside the larger gap opened by Jupiter and Saturn, thermal emission from local concentration of dust at the Lagrangian points of the planets-star system is visible~\citep{Lyra:2009}. 

The assumptions on the disk physical structure are similar as for the models shown in Fig.~\ref{fig:ngvla_planets} but with a factor of $\approx 10\times$ higher gas and dust densities. 
These values are still within the range of the observed properties of disks in nearby star forming regions. In the case of the simulation at 1 cm, the surface brightness of the disk model has been rescaled up to produce a total disk flux of $\approx 1$ mJy at 1 cm, similar to what estimated for the dust emission at this wavelength in a small sample of bright disks in Taurus \citep{Rodmann:2006}. This shows that the ngVLA would identify and characterize structures due to the disk-planet interaction in a Solar Nebula-like disk, at least in the brightest disks known in nearby regions.

The comparison between multi-wavelength ngVLA observations at 3 mm and 1 cm can illuminate differences in the dynamics of small solids due to the different coupling with the gas in the disk. As an example, the ngVLA map at 1 cm shown in Fig.~\ref{fig:ngvla_walsh} displays more pronounced rings than at 3 mm, as a consequence of more efficient dust trapping of larger solids inside a gas ring. Observations of this kind would test models for the dynamics and growth of small solids \citep[e.g.,][]{Pinilla:2012}, which are critical for the understanding of the formation of planetesimals \citep[e.g.,][]{Johansen:2014}.

\subsection{Following the proper motion of disk azimuthal asymmetries}

With the impressive collecting area and angular resolution of the ngVLA, it would be possible to follow the proper motion of the individual azimuthal structures discussed in the previous sections down to orbital radii of $\sim 1$ au. Given the short dynamical timescales at these short separations from the central star, multi-epoch observations with the ngVLA would be able to measure these motions relatively quickly.

At the signal-to-noise ratios of the azimuthal structures shown in the left two panels in Fig.~\ref{fig:ngvla_planets}, the angular precision of phased-referenced
astrometry with the ngVLA would be of about 1 mas, and the proper motion due to the planet orbital rotation
would be detected on a timespan of just a few weeks. A timespan of about 1 week would already be enough to measure the
proper motion of a circumplanetary disk detected with the same signal-to-noise ratio but at an orbital radius of 1 au. 

The detection of the proper motion of these features would be very important to confirm their actual location in the disk, and discard the hypothesis of background or foreground sources. Also, given the short dynamical timescales, one may expect that some of these structures may show some level of dynamical evolution over timescales that the ngVLA will be able to measure. If so, following the time evolution of the flux and morphology of these structures would provide further constraints to theory.

\section{Uniqueness to ngVLA Capabilities}

While the JVLA has frequency coverage up to 50\,GHz, it lacks the necessary sensitivity and angular resolution to detect the physical disk structures discussed here within $\approx 10$ au from the central star. Observations at higher frequencies ($>50$\,GHz) are also important to measure the dust emission towards its brighter fluxes, therefore increasing the signal-to-noise ratio of the detected structures.
In fact, our analysis shows that the 3 mm waveband is really important to be able to characterize the signposts of disk-planet interaction even for an instrument with the collecting area of the ngVLA.

ALMA is significantly more sensitive than the JVLA at detecting disk structures at small scales 
in young disks. 
According to the analysis presented in \citet{Ricci:2018}, ALMA can detect signatures of planets down to $\approx 20 - 30~M_{\oplus}$ at orbital radii $<$ 10 au for the disk parameters considered in that study. However, the gaps opened in these disk regions by planets less massive than Saturn or Jupiter are not well resolved along the radial direction by ALMA. As a consequence, inferring the planet mass would be highly uncertain. Moreover, ALMA cannot detect the multi-ring structures due to Super-Earth planets below $\approx 10 - 20~M_{\oplus}$, as well as several of the azimuthal asymmetries predicted by disk models with low viscosity and which are visible with the ngVLA at 3 mm. Finally, at frequencies below 30 GHz, the Square Kilometre Array (SKA) is not expected to achieve enough sensitivity to study in detail structures generated by planets in disks.






\end{document}